\documentclass[twocolumn,showpacs,preprintnumbers,amsmath,amssymb]{revtex4}
\usepackage{graphicx}

\begin{document}
\draft
\title{Spatio-temporal wave propagation in photonic crystals: a Wannier-function analysis}
\normalsize
\author{Stefano Longhi}
\address{Dipartimento di Fisica, Politecnico di Milano and Istituto di Fotonica e Nanotecnologie del CNR, Piazza L. da
Vinci 32,  I-20133 Milan, Italy}


%
\bigskip
\begin{abstract}
\noindent A general analysis of undistorted propagation of
localized wavepackets in photonic crystals based on a
Wannier-function expansion technique is presented. Different kinds
of propagating and stationary spatio-temporal localized waves are
found from an asymptotic analysis of the Wannier function envelope
equation.
\end{abstract}



\maketitle

\newpage

\maketitle

\section{Introduction}
Spatio-temporal broadening of localized wavepackets with finite
energy due to the effects of diffraction and dispersion is a
universal and challenging phenomenon in any physical context
involving wave propagation. If the finite energy constraint is
left, special spatio-temporal waves with a certain degree of
localization in space and/or in time, capable of propagating free
of diffraction and/or temporal dispersion, can be constructed.
Localized waves of this type include, among others, Bessel beams,
focus-wave modes, X-type waves, pulsed Bessel beams
\cite{Durnin87,Donnelly93,Porras01,Porras03,Longhi03}. Though
these waves can be only approximately realized in practice,
several experiments in acoustic and optical fields have been
reported so far showing nearly-undistorted localized wave
propagation. As the existence of undistorted progressive localized
waves in vacuum has been known since many years and lead to
long-standing studies \cite{Durnin87,Donnelly93}, with special
attention devoted toward their superluminal or subluminal
character and to their finite-energy realizations, in the past few
years these studies have been extended to dispersive optical media
\cite{Porras01,Porras03,Longhi03}, and remarkably the spontaneous
generation of localized and nonspreading wavepackets mediated by
optical nonlinearities has been predicted \cite{Conti03} and
experimentally observed \cite{DiTrapani03} using standard
femtosecond pulsed lasers. Very recently, in a few works
\cite{Conti04,Christodoulides04,Longhi04} the issue of spatial or
spatio-temporal wave localization in {\it periodic} media has been
addressed, and the possibility of exploiting well-established
anomalous diffractive and dispersive properties of photonic
crystals (PCs) \cite{Eisenberg00,Hudock04} to induce novel
spatio-temporal wave localization mechanisms has been proposed.
 Specifically, these studies have been concerned with localization
of Bose-Einstein condensates in a one-dimensional optical lattice
without any trapping potential \cite{Conti04}, with
two-dimensional (2D) spatial Bessel X waves in weakly-coupled 2D
waveguide arrays showing bi-dispersive properties
\cite{Christodoulides04} and with three-dimensional (3D)
out-of-plane X-wave localization in 2D PCs \cite{Longhi04}.
Spatio-temporal waves considered in these works rely on some
specific models and often use ad-hoc approximations, e.g. reduced
coupled-mode equations, paraxiality, weak-coupling limit,
continuum approximations. So far, a general framework to capture
spatio-temporal wave localization and propagation in PCs and the
derivation of a general wave equation, valid regardless of the
specific system under investigation and
with a wide range of applicability, is still lacking.\\
The aim of this work is to provide a general analytical framework
to study spatio-temporal wave propagation in 2D and 3D PCs based
on the use of Wannier-functions, which have been introduced in the
context of PCs to treat localized modes, such as the bound states
of impurities or lattice defects \cite{Albert00,Whittaker03}. A
general asymptotic analysis of the envelope equation for the
Wannier functions allows one to capture the existence and
properties of localized nonspreading wavepackets in PCs in terms
 of localized solutions of canonical wave equations, such as the
 Schr\"{o}dinger equation, the Helmholtz equation and the Klein-Gordon equation.

\section{Wannier function envelope equation}

The starting point of the analysis is provided by the vectorial
wave equation for the magnetic field
$\mathbf{H}=\mathbf{H}(\mathbf{r},t)$ in a PC with a periodic
relative dielectric constant
 $\epsilon(\mathbf{r})$,
\begin{equation}
\nabla \times \left( \frac{1}{\epsilon} \nabla \times
\mathbf{H}\right)=-\frac{1}{c^2} \frac{\partial^2
\mathbf{H}}{\partial t^2},
\end{equation}
where $c$ is the speed of light in vacuum. To study the
propagation of a spatio-temporal wavepacket, we can adopt the
method of the Wannier function expansion, which is commonplace in
the study of the quasi-classical electron dynamics in solids
\cite{Ziman,Feuer52} and recently applied to study localized modes
and defect structures in PCs with defects
\cite{Albert00,Whittaker03}. We refer explicitly to a 3D PC
structure, however a similar analysis can be developed for a 2D
PC. Let us first consider the monochromatic Bloch-type solutions
to Eq.(1) at frequency $\omega$,
$\mathbf{H}(\mathbf{r},t)=\mathbf{H}_{\mathbf{k},n}(\mathbf{r})
\exp(-i \omega t)$, where $\mathbf{k}$ lies in the first Brillouin
zone of the reciprocal $\mathbf{k}$ space,
$\omega=\omega_n(\mathbf{k})$ is the dispersion curve for the
$n$-th band, and $\mathbf{H}_{\mathbf{k},n}(\mathbf{r})$ are the
band modes, satisfying the condition
$\mathbf{H}_{\mathbf{k},n}(\mathbf{r}+\mathbf{R})=\mathbf{H}_{\mathbf{k},n}(\mathbf{r})
\exp(i\mathbf{k}\cdot \mathbf{R})$ for any lattice vector
$\mathbf{R}$ of the periodic dielectric function. The Bloch
functions $\mathbf{H}_{\mathbf{k},n}(\mathbf{r})$ are normalized
such that $\langle \mathbf{H}_{\mathbf{k}',n'} |
\mathbf{H}_{\mathbf{k},n} \rangle= V_{BZ} \; \delta_{n,n'}
\delta(\mathbf{k}'-\mathbf{k}) $, where $V_{BZ}=(2 \pi)^3/V$ is
the volume of the first Brillouin zone in the reciprocal space and
$V$ is the volume of the real-space unit cell. For each band of
the PC, one can construct a Wannier function ${\bf
W}_n(\mathbf{r})$ as a localized superposition of Bloch functions
of the band according to:
\begin{equation}
\mathbf{W}_n(\mathbf{r})=\frac{1}{V_{BZ}} \int_{BZ} d\mathbf{k} \;
\mathbf{H}_{\mathbf{k},n}(\mathbf{r}).
\end{equation}
In the superposition, the phase of Bloch functions
$\mathbf{H}_{\mathbf{k},n}$ can be chosen such that the Wannier
function $\mathbf{W}_n(\mathbf{r})$ is
 strongly localized around $\mathbf{r}=0$ with an
exponential decay far from $\mathbf{r}=0$. The Wannier functions
satisfy the orthogonality conditions $\langle
\mathbf{W}_{n'}(\mathbf{r}-\mathbf{R}')|
\mathbf{W}_{n}(\mathbf{r}-\mathbf{R}) \rangle=\delta_{n,n'}
\delta_{\mathbf{R},\mathbf{R}'}$, and the following relationship
can be easily proven:
\begin{equation}
\langle \mathbf{W}_{n'}(\mathbf{r}-\mathbf{R}')\left| \nabla
\times \left( \frac{1}{\epsilon} \nabla \times \right)
 \right| \mathbf{W}_{n}(\mathbf{r}-\mathbf{R}) \rangle=
 \delta_{n,n'} \theta_{n,\mathbf{R}'-\mathbf{R}} \;\;,
\end{equation}
where $\theta_{n,\mathbf{R}}$ is the Fourier expansion coefficient
of the dispersion curve $\omega_{n}^2 ({\mathbf{k}})$ of the band,
$\theta_{n,\mathbf{R}} \equiv (1/V_{BZ}) \int_{BZ} d\mathbf{k} \;
\omega_{n}^2(\mathbf{k}) \exp(-i\mathbf{k}\cdot\mathbf{R})$, i.e.
$\omega_{n}^2(\mathbf{k})= \sum_{\mathbf{R}} \theta_{n,\mathbf{R}}
\exp(i\mathbf{k}\cdot\mathbf{R})$. We then look for a
spatio-temporal wavepacket, which is a solution to Eq.(1), as a
superposition of translated Wannier functions localized at the
different lattice points $\mathbf{R}$ of the periodic structure,
with amplitudes $f(\mathbf{R},t)$ that depend on the lattice point
$\mathbf{R}$ and can vary in time, i.e. we set:
\begin{equation}
\mathbf{H}(\mathbf{r},t)=\sum_{\mathbf{R}} f(\mathbf{R},t)
\mathbf{W}_n(\mathbf{r}-\mathbf{R}).
\end{equation}
Note that, as we consider a pure periodic structure without
defects and neglect perturbation terms in Eq.(1) (e.g.
nonlinearities), coupling among different bands does not occur and
in Eq.(4) the sum can be taken over a single band, of index $n$.
 Coupled-mode equations for the
temporal evolution of the amplitudes $f(\mathbf{R},t)$ of Wannier
functions at different lattice points can be obtained after
substitution of Eq.(4) into Eq.(1), taking the scalar product with
$\mathbf{W}_n(\mathbf{r}-\mathbf{R})$ and using the orthogonality
conditions of Wannier functions, together with Eq.(3). One
obtains:
\begin{equation}
\frac{\partial^2 f(\mathbf{R},t)}{\partial t^2}+
\sum_{\mathbf{R}'} \theta_{n,\mathbf{R}'-\mathbf{R}}
f(\mathbf{R}',t) =0.
\end{equation}
The solution to the coupled-mode equations (5) can be expressed as
$f(\mathbf{R},t)=f(\mathbf{r}=\mathbf{R},t)$, where the {\it
continuous} function $f(\mathbf{r},t)$ of space $\mathbf{r}$ and
time $t$ satisfies the partial differential equation:
\begin{equation}
\frac{\partial^2 f(\mathbf{r},t)}{\partial t^2}+  \omega_{n}^2(-i
\nabla_{\mathbf{r}} ) f(z,t)=0,
\end{equation}
and $\omega_{n}^2(-i \nabla_{\mathbf{r}})$ is the operator
obtained after the substitution $\mathbf{k} \rightarrow -i
\nabla_{\mathbf{r}}$ in the Fourier expansion of
$\omega_{n}^2(\mathbf{k})$. It should be noted that the
differential equation for the {\it continuous envelope}
$f(\mathbf{r},t)$ of the Wannier function wavepacket [Eq.(4)], as
given by Eq.(6), is exact, and for any band of the PC an envelope
equation can be written, the specific details of the band entering
both in the dispersion curve $\omega_{n}^2(\mathbf{k})$ and in the
shape of the corresponding Wannier function $\mathbf{W}_n$
[Eq.(2)].

\section{Spatial and spatio-temporal localized waves}
 The most general solution to the Wannier-function
envelope equation (6) is given by a superposition of functions
$\psi(\mathbf{r}, \pm t)$, where $\psi(\mathbf{r},t)$ is a
solution to the wave equation:
\begin{equation}
i \frac{\partial \psi}{\partial t}= \omega_n(-i
\nabla_{\mathbf{r}}) \psi.
\end{equation}
We are know interested on the search for localized solutions to
Eq.(7) such that $|\psi|$ corresponds to a wave propagating
undistorted with a group velocity $v_g$. To this aim, let us set
$\psi(\mathbf{r},t)=g(\mathbf{r},t) \exp(i\mathbf{k}_0-i \Omega
t)$, where $\mathbf{k}_0$ is chosen inside the first Brillouin
zone in the reciprocal space and the frequency $\Omega$ is chosen
close to (but not necessarily coincident with)
$\omega_0=\omega_n(\mathbf{k}_0)$. The envelope $g$ then satisfies
the wave equation
\begin{equation}
i  \frac{\partial g}{
\partial t}= \left [\omega_n(\mathbf{k}_0-i \nabla_{\mathbf{r}})-\Omega \right]g.
\end{equation}
We first note that, if $g$ varies slowly with respect to the
spatial variables $\mathbf{r}$, at leading order one can expand
$\omega_n(\mathbf{\mathbf{k}}_0-i \nabla_{\mathbf{r}})$ up to
first order around $\mathbf{k}_0$; taking $\Omega=\omega_0$, one
obtains $ \partial g / \partial t+ \nabla_{\mathbf{k}} \omega_n
\cdot \nabla_\mathbf{r} g=0$, i.e. one retrieves the well-known
result for which {\it an arbitrary} 3D spatially-localized
wavepacket travels undistorted, at leading order, with a group
velocity given by $\nabla_{\mathbf{k}} \omega_n$. Nevertheless,
higher-order terms are generally responsible for wavepacket
spreading, both in space and time. In order to find
propagation-invariant envelope waves {\it even when dispersive
terms} are accounted for, let us assume, without loss of
generality, that $(\partial \omega_n /
\partial k_y)_{\mathbf{k}_0}=(\partial \omega_n / \partial
k_z)_{\mathbf{k}_0}=0$, i.e. let us choose the orientation of the
$x$ axis such that the wavepacket group velocity
$\nabla_{\mathbf{k}} \omega_n$ is directed along this axis, and
let us look for a propagation-invariant solution to Eq.(8) of the
form $g=g(x_1,x_2,x_3)$, with $x_1=x-v_gt$, $x_2=y$ and $x_3=z$,
traveling along the $x$ axis with a group velocity $v_g$, which is
left undetermined at this stage. The function $g$ then satisfies
the following equation:
\begin{equation}
-iv_g \frac{\partial g}{\partial x_1}=
\left[\omega_n(\mathbf{k}_0-i \nabla_{\mathbf{x}}) -\Omega
\right]g,
\end{equation}
whose solution can be written formally as:
\begin{equation}
g(x_1,x_2,x_3)= \int dQ_2 dQ_3 \; G(Q_2,Q_3) \exp (i \mathbf{Q}
\cdot \mathbf{x}).
\end{equation}
In Eq.(10), $\mathbf{x}=(x_1=x-v_gt,x_2=y,x_3=z)$,
$\mathbf{Q}=(Q_1,Q_2,Q_3)$, $G$ is an arbitrary spectral
amplitude, and $Q_1=Q_1(Q_2,Q_3)$ is implicitly defined by the
following {\it dispersion relation}:
\begin{equation}
\omega_n(\mathbf{k}_0+\mathbf{Q})-\Omega-v_g Q_1=0.
\end{equation}
To avoid the occurrence of evanescent (exponentially-growing)
waves, the integral in Eq.(10) is extended over the values of
$(Q_2,Q_3)$ such that $Q_1$, obtained after solving Eq.(11), turns
out to be real-valued. We note that, for an {\it arbitrary}
spectral amplitude $G$, Eq.(10) represents an {\it exact} solution
of the Wannier-function envelope equation, which propagates {\it
undistorted} with a group velocity $v_g$, once the proper band
dispersion curve $\omega_n(\mathbf{k})$ of the PC and
corresponding dispersion relation (11) are computed, e.g. by
numerical methods. For {\it some} specific choices of the spectral
amplitude $G$, in addition to undistorted wave propagation a
certain degree of spatio-temporal wave localization can be
obtained. It is worth to get some explicit examples, though
approximate, of such 3D localized waves, admitting the integral
representation given by Eq.(10), and relate them to already known
localized solutions to canonical wave equations \cite{Donnelly93}.
To this aim, we develop an asymptotic analysis of Eq.(11) by
assuming that the spectral amplitude $G$ is nonvanishing in a
narrow interval around $Q_2=Q_3=0$, so that, for $\Omega$ close to
$\omega_0$, the value of $Q_1$, as obtained form Eq.(11), is also
close to $Q_1=0$. In this case, an approximate expression for the
dispersion relation $Q_1=Q_1(Q_2,Q_3)$ can be obtained by
expanding in Eq.(11) the band dispersion curve
$\omega_n(\mathbf{k}_0+\mathbf{Q})$ at around $\mathbf{k}_0$. We
should distinguish two cases, depending on the value of the group
velocity $v_g$, which is basically a free parameter in our analysis. \\
{\it First case}. The first case corresponds to the choice of a
group velocity $v_g$ different from (and enough far form)
$\partial \omega_n /
\partial k_x$. In this case, the leading-order terms entering in Eq.(11)
after a power expansion of $\omega_n(\mathbf{k}_0+\mathbf{Q})$ are
quadratic in $Q_2$, $Q_3$ and linear in $Q_1$; precisely, one has:
\begin{equation}
\left( \frac{\partial \omega_n}{\partial k_1}-v_g
\right)Q_1+\omega_0-\Omega+\frac{1}{2} \sum_{i,j=2}^{3}
\frac{\partial^2 \omega_n}{\partial k_i \partial k_j} Q_i Q_j=0,
\end{equation}
where $k_i=k_{x,y,z}$ for $i=1,2,3$ and the derivatives of the
band dispersion curve are calculated at $\mathbf{k}=\mathbf{k}_0$.
If the approximate expression of $Q_1$, given Eq.(12), is
introduced into Eq.(10), one can easily show that the envelope
$g(x_1,x_2,x_3)$ satisfies the differential equation:
\begin{equation}
i \left( \frac{\partial \omega_n}{\partial k_1}-v_g
\right)\frac{\partial g}{\partial
x_1}=(\omega_0-\Omega)g-\frac{1}{2} \sum_{i,j=2}^3
\frac{\partial^2 \omega_n}{\partial k_i
\partial k_j} \frac{\partial^2 g}{\partial x_i \partial x_j}.
\end{equation}
Since the matrix $\partial^2 \omega_n / \partial k_i \partial k_j$
is symmetric, after a suitable rotation of the $(x_2,x_3)$ axes by
the transformation $x^{'}_j=\mathcal{R}_{ji}x_i$ ($i,j=2,3$),
where $\mathcal{R}_{ji}$ is the orthogonal matrix that
diagonalizes $\partial^2 \omega_n / \partial k_i \partial k_j$,
assuming without loss of generality $\Omega=\omega_0$, Eq.(13) can
be written in the canonical Schr\"{o}dinger-like form:
\begin{equation}
i \left( \frac{\partial \omega_n}{\partial k_1}-v_g
\right)\frac{\partial g}{\partial x_1}=-\frac{1}{2} \alpha_2
\frac{\partial^2 g}{\partial x^{'2}_{2} }--\frac{1}{2} \alpha_3
\frac{\partial^2 g}{\partial x^{'3}_{3} },
\end{equation}
where $\alpha_2$ and $\alpha_3$ are the eigenvalues of the $2
\times 2$ matrix $\partial^2 \omega_n / \partial k_i \partial k_j$
($i,j=2,3$). 3D localized waves to Eq.(14) are expressed in terms
of well-known Gauss-Hermite functions, which are in general
anisotropic for $\alpha_2 \neq \alpha_3$. These 3D localized
waves, which exist regardless of the sign of $\alpha_2$ and
$\alpha_3$, represent Gaussian-like beams, with exponential
localization in the transverse $(y,z)$ plane and algebraic
localization, determined by the beam Rayleigh range, in the
longitudinal $x$ direction (and hence in time). These beams
propagate undistorted along the $x$ direction with an {\it
arbitrary} group velocity $v_g$, either subluminal or
superluminal, provided that $v_g \neq \partial \omega_n /
\partial k_x$. Such pulsed propagating Gaussian
beams represent an extension, in a PC structure, of similar
solutions found in vacuum (see \cite{Longhi04b} and references
therein). In particular, the special case $v_g=0$ leads to
stationary (monochromatic) Gaussian-like beams; note that the
condition $v_g \neq \partial \omega_n / \partial k_x$ implies that
such steady Gaussian beams do not exist in a PC close to a bandgap
edge, where $\partial \omega_n /
\partial k_x$ vanishes. Other solutions to Eq.(14), leading to {\it
spatial} 2D localized and monochromatic waves in the transverse
$(y,z)$ plane (but delocalized in the longitudinal $x$ direction),
can be search in the form $g(x_1,x_2,x_3)=s(x_2,x_3) \exp(i
\lambda x_1)$, where $\lambda$ is a propagation constant. If
$\alpha_2$ and $\alpha_3$ have the same sign, the function
$s(x_2,x_3)$ satisfies a 2D Helmholtz equation, admitting
well-known Bessel-beam solutions in cylindrical coordinates. For
$\alpha_2 \neq \alpha_3$, such solutions are anisotropic, and
again they represent a generalization to a PC of well-known
spatial Bessel beams in vacuum. If $\alpha_2$ and $\alpha_3$ have
opposite sign, one obtains a hyperbolic 2D equation (or,
equivalently, a 1D Klein-Gordon equation), which admits of 2D
X-type localized solutions involving modified Bessel functions
recently studied in \cite{Christodoulides04} (see Eqs.(3a) and (4)
of Ref. \cite{Christodoulides04}; see also \cite{Ciattoni04}).\\
{\it Second case.} The second case corresponds to the choice
$v_g=\partial \omega_n / \partial k_x$. In this case, the
leading-order approximation to the dispersion relation [Eq.(11)]
should include also second-order derivatives  with respect to
$x_1$ of the band dispersion curve
$\omega_n(\mathbf{k}_0+\mathbf{Q})$, yielding:
\begin{equation}
\omega_0-\Omega+\frac{1}{2} \sum_{i,j=1}^3 \frac{\partial^2
\omega_n}{\partial k_i \partial k_j} Q_i Q_j=0,
\end{equation}
where the derivatives of the band dispersion curve are calculated
at $\mathbf{k}=\mathbf{k}_0$. If the approximate expression of
$Q_1$, implicitly defined by the quadratic equation (15), is
introduced into Eq.(10), one can easily show that the envelope
$g(x_1,x_2,x_3)$ satisfies this time the differential equation:
\begin{equation}
(\omega_0-\Omega)g=\frac{1}{2} \sum_{i,j=1}^3 \frac{\partial^2
\omega_n}{\partial k_i \partial k_j} \frac{\partial^2 g}{\partial
x_i \partial x_j}.
\end{equation}
Since the matrix $\partial^2 \omega_n / \partial k_i \partial k_j$
is symmetric, after a suitable rotation of the $(x_1,x_2,x_3)$
axes by the transformation $x^{'}_j=\mathcal{R}_{ji}x_i$
($i,j=1,2,3$), where $\mathcal{R}_{ji}$ is the orthogonal matrix
that diagonalizes $\partial^2 \omega_n / \partial k_i \partial
k_j$, Eq.(16) takes the canonical form:
\begin{equation}
(\omega_0-\Omega)g=\frac{1}{2} \left( \alpha_1 \frac{\partial^2
g}{\partial x^{'2}_{1} } + \alpha_2 \frac{\partial^2 g}{\partial
x^{'2}_{2} }+ \alpha_3 \frac{\partial^2 g}{\partial x^{'3}_{3} }
\right),
\end{equation}
where $\alpha_i$ ($i=1,2,3$) are the eigenvalues of the $3 \times
3$ matrix $\partial^2 \omega_n / \partial k_i \partial k_j$
($i,j=1,2,3$). The sign of the eigenvalues $\alpha_i$ basically
determines the elliptic or hyperbolic character of Eq.(17), and
hence the nature of their solutions (see, e.g.,
\cite{Donnelly93}). If $\alpha_i$ have the same sign, e.g. they
are positive, for $\Omega<\omega_0$ Eq.(17) reduces, after a
scaling of axis length, to a 3D Helmholtz equation, which in
spherical coordinates admits of localized solutions in the form of
sinc-shaped waves (see, e.g., \cite{Donnelly93,Longhi03}). If,
conversely, there is a sign discordance among the eigenvalues
$\alpha_i$, one obtains a 2D Klein-Gordon equation, which admits
of 3D localized X-type waves which have been lengthly discussed in
many works (see, e.g., \cite{Donnelly93,Conti03,Christodoulides04}
and references therein). In some special cases, one of the
eigenvalues $\alpha_i$ may vanish, which may yield further
nonspreading wavepacket solutions. Notably, if $\alpha_1=0$, the
solution to Eq.(17) is given by $g(x_1,x_2,x_3)=h(x_1)
\varphi(x_2,x_3)$, where $h$ is an arbitrary function of
$x_1=x-v_gt$ and $\varphi$ satisfies a 2D Helmoltz equation for
$\alpha_2 \alpha_3>0$, admitting Bessel beam solutions, or a 1D
Klein-Gordon equation for $\alpha_2 \alpha_3 <0$, admitting 2D
X-type solutions. For these special solutions a cancellation of
temporal dispersion is attained. As the former case ($\alpha_2
\alpha_3>0$) extends to a PC structure the so-called pulsed Bessel
beams found in homogeneous dispersive media \cite{Porras01}, the
latter case ($\alpha_2 \alpha_3 <0 $) is rather peculiar for a PC
structure, which realizes a bi-diffractive propagation regime
\cite{Christodoulides04}, i.e. positive and negative diffraction
along the two transverse directions $y$ and $z$. Instead of pulses
with a transverse Bessel beam profile, in this case one obtains a
transverse X-shaped beam
with an arbitrary longitudinal (temporal) profile that propagates without spreading.\\
As a final remark, we note that, though our analysis has been
focused to a 3D PC, similar results can be obtained {\it mutatis
mutandis} for the lower-dimensional case of a 2D PC. In this case,
not considering out-of-plane propagation, the fields depend solely
on the two spatial variables $x$ and $y$ defining the PC plane,
and Eqs.(14) and (17) are still valid provided that the terms
involving the derivatives with respect to the $x_3=z$ coordinate
are dropped. In this case, Eq.(14) corresponds to a 1D
Schr\"{o}dinger equation, whereas Eq.(17) corresponds to either a
2D Helmholtz equation or to a 1D Klein-Gordon equation.

\section{Conclusions}
In conclusion, a general analysis of wavepacket
propagation in PCs, based on a Wannier function expansion
approach, has been presented, and an exact envelope equation
describing undistorted propagation of spatio-temporal localized
waves has been derived. An asymptotic analysis of the envelope
equation shows that a wide class of localized (either spatial or
spatio-temporal) waves exist, including propagating Gaussian
beams, 2D and 3D X-type waves, sinc-shaped waves, pulsed Bessel
beams and pulsed 2D X waves, some of which have been recently
studied with reference to some specific models
\cite{Conti04,Christodoulides04}.

\end{document}